\documentclass[english,prd,superscriptaddress,nofootinbib,
               preprintnumbers,twocolumn,showpacs]{revtex4}

\usepackage[latin1]{inputenc}
\usepackage{amsmath}
\usepackage{amssymb}
\usepackage{graphicx}
\usepackage[caption=false]{subfig}
\usepackage{amsthm}
\usepackage{bm}
\usepackage{color}
\usepackage{amsfonts}
\usepackage{dcolumn}
\usepackage{babel}

\def\fp{f^{\prime}}
\def\fpp{f^{\prime \prime}}

\makeatletter

\def\be{\begin{equation}}
\def\ee{\end{equation}}
\def\ba{\begin{eqnarray}}
\def\ea{\end{eqnarray}}

\begin{document}
\title{On the absence of the usual weak-field limit, and the impossibility of embedding some known solutions for isolated masses in cosmologies with f(R) dark energy}
\author{Timothy Clifton}
\affiliation{School of Physics and Astronomy,
 Queen Mary University of London, 
Mile End Road, London E1 4NS, UK.}

\author{Peter Dunsby}
\affiliation{Astrophysics Cosmology \& Gravity Center, and Department of Mathematics \& Applied Mathematics,
University of Cape Town, 7701 Rondebosch, South Africa.}
\affiliation{South African Astronomical Observatory,  Observatory 7925, Cape Town, South Africa.}

\author{Rituparno Goswami}
\affiliation{Astrophysics Cosmology \& Gravity Center, and Department of Mathematics \& Applied Mathematics,
University of Cape Town, 7701 Rondebosch, South Africa.}

\author{Anne Marie Nzioki}
\affiliation{Astrophysics Cosmology \& Gravity Center, and Department of Mathematics \& Applied Mathematics,
University of Cape Town, 7701 Rondebosch, South Africa.}
\begin{abstract}
The problem of matching different regions of spacetime in order to construct inhomogeneous cosmological models is investigated in the context of Lagrangian theories of gravity constructed from general analytic functions $f(R)$, and from non-analytic theories with $f(R)=R^n$. In all of the cases studied, we find that it is impossible to satisfy the required junction conditions without the large-scale behaviour reducing to that expected from Einstein's equations with a cosmological constant. For theories with analytic $f(R)$ this suggests that the usual treatment of weak-field systems as perturbations about Minkowski space may not be compatible with late-time acceleration driven by anything other than a constant term of the form $f(0)$, which acts like a cosmological constant. In the absence of Minkowski space as a suitable background for weak-field systems, one must then choose and justify some other solution to perform perturbative analyses around. For theories with $f(R)=R^n$ we find that no known spherically symmetric vacuum solutions can be matched to an expanding FLRW background. This includes the absence of any Einstein-Straus-like embeddings of the Schwarzschild exterior solution in FLRW spacetimes.
\end{abstract}
\maketitle
\section{Introduction}
Fourth-order theories of gravity have recently attracted a considerable amount of attention as they admit Friedmann-Lema\^{i}tre-Robertson-Walker (FLRW) solutions that can accelerate at late times without the presence of any exotic fluids. It then follows that the apparent need for dark energy could simply be due to an inappropriate application of  Einstein's equations to scales beyond those within which they have been thoroughly tested. This idea is compelling, but constitutes a radical shift from the standard approach to cosmology. It must therefore be carefully studied in order to ensure that the consequences of deviating away from Einstein's theory are fully understood. In this paper we attempt to contribute to this understanding by considering the construction of inhomogeneous cosmological models within the framework of $f(R)$ theories of gravity.

The gravitational fields around isolated objects, and the FLRW solutions of $f(R)$ theories, have both been extensively studied in the literature (see \cite{fR1,fR2,fR3,fR4,MG} for reviews). Here we do not intend to contribute further to the study of either of these fields, but instead to the ways in which one can construct cosmological models that contain massive astrophysical bodies. This will be done by attempting to match together existing solutions. In particular, we will attempt to construct `Swiss cheese' models by matching spherically symmetric vacuum solutions with FLRW solutions, as well as constructing `lattice models' by matching together large numbers of the spacetimes associated with regularly spaced astrophysical bodies.

The motivation for this study is to understand both the effect of cosmological expansion on the gravitational fields of astrophysical bodies, as well as the large-scale expansion that emerges in a universe with large density contrasts. These questions have been carefully studied in Einstein's theory, where the above constructions have proven to be useful devices for understanding them. Fourth-order theories are considerably more complicated than Einstein's theory, but by applying the same constructions we should expect to gain some insights into these questions. These extra complications include the absence of Birkhoff's theorem, so that spherically symmetric vacuum spacetimes are not unique \cite{sph}, as well as more complicated junction conditions \cite{der}.

Further motivation for this study comes specifically from the work of Mignemi and Wiltshire \cite{mig}. These authors used a dynamical systems approach to perform a non-perturbative study of the static, spherically symmetric solutions of analytic $f(R)$ theories. They found that these solutions are generically not asymptotically flat, and that boundary conditions could therefore be important in determining the gravitational fields of isolated massive bodies. Similar results have been found for non-analytic $f(R)$ theories \cite{power}. These effects are entirely absent if one {\it assumes} asymptotic flatness from the beginning, as is standard in most approaches to studying weak gravitational fields \cite{pech,fRPPN}. The construction of inhomogeneous cosmological models, as outlined above, provides a way to implement appropriate boundary conditions, and therefore allows the validity of standard weak-field approaches to be investigated.

At present, much of the current literature assumes that in $f(R)$ theories the evolution of the FLRW `background' cosmology proceeds independently of the growth of structure within it. The motivation for this within Einstein's theory comes, in large part, from the studies of inhomogeneous cosmologies, such as those discussed above. It also comes, however, from the correspondence between Newtonian cosmology and the FLRW solutions of Einstein's equations during dust domination: The rate at which nearby astrophysical bodies fall away from each other can be considered as being due to a Newtonian force (up to the usual accuracy this implies), or due to the expansion of the universe. Both are reasonable descriptions on small enough scales. If one attempts to use $f(R)$ as an explanation of dark energy, however, then one wants the cosmological expansion to be {\it different} to that of a dust dominated universe. The usual interpretation of the motion of nearby astrophysical bodies as being describable (up to some accuracy) within Newtonian theory is therefore lost, and the intuition we have gained on this subject from studying the solutions of Einstein's equations must be re-evaluated.

A thorough investigation of this problem is a very difficult proposition, as in order to evaluate the existence or not of a weak-field limit, and the emergence of FLRW-like behaviour on large scales, one cannot begin by assuming the existence of either of these things. Any realistic investigation, however, needs to make some assumptions, and here we will begin by assuming that the gravitational fields around astrophysical bodies can be described by known solutions (either weak-field or exact). We will then proceed to see which FLRW solutions these can be matched with, or which FLRW behaviours emerge, given this assumption. This approach will not allow all of the questions posed above to be answered fully, but will allow us to show that some situations that are possible in Einstein's theory are not possible in all $f(R)$ theories.

In Section \ref{sec:fr} we introduce $f(R)$ theories of gravity, and in Section \ref{sec:weak} we discuss what we mean by `the weak-field limit'. This includes taking Minkowski space to be the solution around which weak-field expansions are performed. Then in Section \ref{sec:junction} we discuss the junction conditions that need to be satisfied when matching together different solutions in $f(R)$ theories. In Section \ref{sec:swiss} we attempt to make a Swiss-cheese-like construction in which we match the usual weak-field solutions to FLRW solutions in theories with analytic $f(R)$. Section \ref{sec:lattice} contains an attempt within these same theories to match together many different weak-field regions to make a lattice model of the universe. In both Sections \ref{sec:swiss} and \ref{sec:lattice} it is found that these constructions are only possible if the large-scale behaviour is similar to that of solutions to Einstein's equations with a cosmological constant. In Section \ref{sec:exact} we then proceed to try and match some known exact solutions, including here some theories with non-analytic $f(R)$. We take three known exact solutions that describe spherically symmetric vacua in these theories, and try and match each to FLRW solutions. We find that in no situations can the junction conditions be satisfied at the boundaries between regions. Finally, in Section \ref{sec:conclude} we discuss our results.

Throughout this paper we will us Greek letters to denote spacetime indices, and Latin letters $a$, $b$, $c$ {\it etc.} to denote coordinates on a boundary. When it is required, the letters $i$, $j$, $k$ {\it etc.} will be reserved for spatial indices.

\section{$f(R)$ theories of gravity}
\label{sec:fr}
The action for $f(R)$ theories is given by replacing the Ricci scalar, $R$, in the Einstein-Hilbert action by a function of the Ricci scalar, $f(R)$, so that the gravitational Lagrangian density is
$\mathcal{L} = f(R)$.

Including a matter action, and extremizing with respect to the metric, the field equations for these theories can then be written as
\be
G_{\mu\nu}
=\frac{T_{\mu\nu}^m}{f'}+T_{\mu\nu}^R\,,
\label{FE}
\ee
where 
\be
T_{\mu\nu}^R=\frac{1}{f'}\left[ \frac12 (f-Rf') g_{\mu\nu}
+ \nabla_{\nu}\nabla_{\mu}f'- g_{\mu\nu}\nabla_{\alpha}\nabla^{\alpha}f' \right]
\label{FE2}
\ee
is the effective energy-momentum tensor of what we will call the ``curvature fluid'', and $T^m_{\mu \nu}$ is the energy-momentum tensor of the standard matter fields. Primes denote differentiation with respect to $R$.

That the field equations (\ref{FE}) are fourth order in derivatives of the metric can be seen from the existence of the $\nabla_{\nu}\nabla_{\mu}f'$  term in (\ref{FE2}), a result which also follows directly from Lovelock's theorem.  This is generally thought of as an undesirable feature in a Lagrangian based theory as it can lead to Ostrogradski instabilities in the solutions of the field equations.  The $f(R)$ theories, however, are a special case in which this instability can be avoided \cite{Woodard}, due to the existence of an equivalence with scalar-tensor theories.  In the special case $f(R) = R$ it can be seen that the fourth-order terms vanish, and Einstein's equations are recovered.

The field equations (\ref{FE}) are automatically generally covariant and Lorentz invariant as they are derived from a Lagrangian that is a function of $R$ only, and $R$ is a generally covariant and locally Lorentz invariant scalar quantity.  These same symmetries also guarantee that the left-hand side of (\ref{FE}) is covariantly conserved.  The $f(R)$ theories therefore exhibit many of the key features of general relativity, while generalising Einstein's equations to allow new behaviour.  This freedom has been shown to allow improved renormalisation of the gravitational interaction \cite{renorm}, as well as early universe inflation \cite{inflation}, and a possible explanation of dark energy.
\section{The Weak-Field Limit}
\label{sec:weak}
Before progressing further, let us begin by specifying exactly what we mean by the term `weak-field limit'. We take this phrase to mean that in extended regions of the Universe that are small compared to the Hubble scale, but large compared to the Schwarzschild radius of any compact objects that may exist within it, that the geometry of spacetime within the region (but outside of the compact objects) can be well described by small fluctuations around Minkowski space, such that
\be
\label{weakfield}
g_{\mu \nu} \simeq \eta_{\mu \nu} + h_{\mu \nu}\;,
\ee
where $\eta_{\mu \nu}$ is the metric of Minkowski space, and there exists a coordinate system in which each of the components of $h_{\mu \nu}$ is $\ll 1$ and slowly varying.  The description given by Eq. (\ref{weakfield}), and the corresponding physics, is what is meant by `the weak-field limit'.

There are a number of points in this explanation that require further clarification.  Firstly, what we mean by `Hubble scale' here is the quantity $cH^{-1}$ when considering space-like separations, and $H^{-1}$ when considering time-like separations (here $H$ is the Hubble constant, as measured by observers using the recessional velocity of nearby objects).  For a region to be `small' compared to the Hubble scale then means that any two points on the boundary of that region that are space-like separated should be $\ll cH^{-1}$ apart, and that any two points that are time-like separated should be $\ll H^{-1}$ apart.  This definition requires $H$ to be reasonably uniform throughout each small region, which we will assume to be true.  The criterion that these regions should be much larger than the Schwarzschild radius of any compact objects, and that Eq. (\ref{weakfield}) should not be taken to describe the regions inside (or near) compact objects, are simply intended to remove from our consideration the regions near black holes and neutron stars.

Let us now further consider Eq. (\ref{weakfield}). The crucial point here is that the geometry of spacetime in the region under consideration can be taken to be close to that of Minkowski space. In this case one can decompose the tensor $h_{\mu \nu}$ according to how its various parts transform under spatial rotations in the background Minkowski space.  In general, one can then write $h_{\mu \nu}$ as (see \cite{Bardeen})
\be
\nonumber
h_{\mu \nu} dx^{\mu} dx^{\nu} = 
2 \phi c^2 dt^2 
-2 B_i c dt dx^i 
+2 \left( \psi \delta_{ij} + H_{ij} \right) dx^i dx^j.
\ee
The divergence of $B_i$ and the trace of $H_{ij}$ can be set to zero by an appropriate choice of coordinates, and the divergence-less part of $B_i$ and the trace-free part of $H_{ij}$ can be consistently ignored. This leaves
\be
\label{Newtonian}
ds^2 \simeq -(1-2 \phi) c^2 dt^2 + (1+2 \psi) \delta_{ij} dx^i dx^j,
\ee
where $\phi$ and $\psi$ are both $\ll 1$ and slowly varying.  We refer to this as `the Newtonian limit' if $\phi$ behaves like a Newtonian potential, and satisfies $\nabla^2 \phi \simeq -4 \pi G \rho$.

Finally, we can make the concepts of `small' and `slow' precise by introducing a dimensionless order-of-smallness parameter, $\epsilon$.  Velocities, $v^i=dx^i/dt$, are then said to be `small' if $v/c \sim O(\epsilon)$, and quantities are said to be `slowly varying' if acting on them with a time-derivative adds an extra $O(\epsilon)$ of smallness when compared to a spatial derivative (the order of smallness of time derivatives and velocities are expected to be similar because the evolution of gravitating systems are typically governed by the motion of their constituents).  From the field equations and equations of motion it can then be seen that the lowest order parts of $\phi$ and $\psi$, and the energy density $\rho$, are given by
\be
\nonumber
\phi \sim \psi \sim G \rho \sim \frac{v^2}{c^2} \sim \epsilon^2.
\ee
The field equations and equations of motion within the region under consideration can then be expanded order-by-order in $\epsilon$, with the `weak field' limit of Eq. (\ref{Newtonian}) corresponding to the expansion up to $O(\epsilon^2)$.  The $\simeq$ sign will be used in what follows to mean `equal up to terms of $O(\epsilon^3)$ and smaller'.  This is the same expansion in $\epsilon$ that is routinely used in the standard parameterised post-Newtonian approach to gravitational physics in weak-field systems \cite{Will}.

One may note the differences between the perturbative expansion outlined here, and the one that is routinely used in cosmological perturbation theory about an FLRW background.  This difference is intentional, and indeed necessary, for the study we are performing. Cosmological perturbation theory is designed to be used on a variety of scales, all the way up to the scale of the cosmological horizon.  This necessitates using a background that is {\it not} Minkowski space and fields that are {\it not} slowly varying, as trying to model such a situation with a Minkowski background would involve 3-velocities that are of the same order of magnitude as the speed of light.  Instead, here we are only interesting in modelling regions that are much smaller than the horizon size using our perturbative framework, and then building up a cosmological model by joining together many of these regions. In this case the order-of-smallness parameter, $\epsilon$, can thought of (approximately) as the scale of the small region compared to the Hubble scale of the eventual cosmological spacetime. In small regions such as this it is well known that relevant quantities become slowly varying, and it is this slowness that is formally incorporated into the perturbative expansion in weak-field systems by allowing time derivatives to add an extra power of $\epsilon$ to a quantity.

The weak-field limit of $f(R)$ theories of gravity has been studied extensively in the literature (see e.g. \cite{fR1,fR2,fR3,fR4,MG}, and references therein), with the full post-Newtonian limit of theories with analytic $f(R)$ that admit Minkowski space as a solution being found in \cite{fRPPN}.  There the Lagrangian function is expanded in a Taylor series as
\be
f(R) = f(0) + \fp(0) R + \frac{1}{2} \fpp(0) R^2 + O(R^3),
\ee
where primes denote differentiation with respect to $R$. One may note here that the expansion is being performed as a series around $R=0$, in keeping with our assumption that Minkowski space is a suitable background about which we can perform an analysis of the weak field. This limits our consideration to theories in which $f(0)$, $\fp(0)$ {\it etc.} are finite numbers, which is certainly not true for all theories (see, e.g., \cite{rinv}). One is, of course, at liberty to consider expanding around other backgrounds, with non-zero Ricci curvature, $R_0$ (see, e.g., \cite{olmo}). In this case, however, one must deal with the complexity of solving the full non-linear Einstein equations in order to find the background, which is both difficult and likely to result in multiple different possibilities. We will consider this further for some simple theories in Section \ref{sec:exact}.

To the order required here, and taking Minkowski space as the background geometry, the metric is given by Eq. (\ref{Newtonian}) with \cite{fRPPN}
\ba
\phi &=& \frac{1}{\fp_0} \left( \hat{U} + \frac{1}{2} \fpp_0 R \right)\;,\\
\psi &=& \frac{1}{\fp_0} \left( U - \frac{1}{2} \fpp_0 R \right)\;,
\ea
where we have used the abbreviations $\fp_0=\fp (0)$ and $\fpp_0=\fpp (0)$, and where $U$, $\hat{U}$ and $R$ satisfy
\be
\label{PPNU}
\nabla^2 U = -4 \pi \rho +\frac{f_0}{4}, \qquad
\nabla^2 \hat{U} = -4 \pi \rho -\frac{f_0}{2} 
\ee
and
\be
\label{PPNR}
\nabla^2 R - \frac{\fp_0}{3 \fpp_0} R = - \frac{8 \pi}{3 \fpp_0} \rho + \frac{2f_0}{3f^{\prime\prime}_0},
\ee
where $f_0=f(0)$. In these equations, and in what follows, we have chosen to use geometrised units in which $G=1=c$.  

Assuming the existence of a weak-field limit, these theories can be seen to have a Newtonian limit if $\fpp R \ll U$.  Unlike in the PPN treatment, we will not insist that the solutions of Eqs. (\ref{PPNU}) and (\ref{PPNR}) approach zero at asymptotically large distances, but will instead enforce boundary conditions using cosmology.
\section{Junction Conditions}
\label{sec:junction}
In what follows we will be matching together different regions of spacetime in order to construct inhomogeneous cosmological models.  This requires a set of junction conditions, analogous to the Israel junction conditions from general relativity \cite{Israel}, and is a problem that has been considered in $f(R)$ theories of gravity by Deruelle, Sasaki and Sendouda \cite{der}.  We will briefly recap the relevant results from their work here, as it of central importance to our study.

The central requirement in \cite{der} is that if one allows delta functions in the matter part of the field equations (i.e. if one allows matter fields to be localised on the boundary hyper-surface), then delta functions should occur at most linearly in the parts of the field equations that involve geometry only. Here we are interested in the case in which there is no brane located at the boundary. We therefore require that there should be {\it no} delta functions in the part of the field equations containing geometry only.

Now, in a Gaussian normal coordinate system, $ds^2 = dy^2 + \gamma_{a b} dx^{a} dx^{b}$, where the boundary is located at $y=0$, the Ricci scalar can be written as
\be
\label{R}
R = 2 \partial_y K - K^*_{ab}K^{*ab} - \frac{4}{3} K^2 + \bar{R},
\ee
where $\bar{R}$ is the Ricci curvature constructed from $\gamma_{ab}$, the extrinsic curvature of the boundary is $K_{ab}= - \frac{1}{2} \partial_y \gamma_{ab}$, and $K$ and $K^{*}_{ab}$ are the trace and trace-free parts of this quantity, respectively.

It can be seen from the field equations (\ref{FE}) that $R$ must be continuous at the boundary. This is because the curvature fluid contains terms like $\partial_{y} \fp(R)$, which can be expanded as
\be
\label{Rjunc}
\partial_{y} \fp(R) = \fpp(R) \partial_{y} R.
\ee
If $R$ is not continuous then the second term above would contain a factor of $\delta(y)$, this is not allowed unless $f^{\prime \prime}(R)=0$, which is just Einstein's equations. We can then see from Eq. (\ref{R}) that $\gamma_{ab}$ must also be continuous, otherwise $K_{ab}$ would contain a factor of $\delta(y)$, and $R$ would contain factors of $(\delta(y))^2$. This is not allowed, as $K_{ab}$ and $R$ occur directly in the field equations. We therefore have that $\gamma_{ab}$ and $R$ must both be continuous across the boundary.

The $yy$ and $ya$ components of Eq. (\ref{FE}) are then given by
\be
\partial_y \left[ \left(K_{ab} -K \gamma_{ab}\right) \fp(R) + \gamma_{ab} \fpp(R) \partial_y R \right] = 0.
\ee
Integrating this across the boundary one then finds
\be
\label{bc1}
\left[ \left(K_{ab} -K \gamma_{ab}\right) \fp(R)+ \gamma_{ab} \fpp(R) \partial_y R \right]^+_-=0,
\ee
where the $[ \dots ]^+_-$ notation means the difference of the quantity in the brackets on either side of the boundary. Similarly, one can integrate $R$ across the boundary to find, from Eq. (\ref{R}) that $[ R ]^+_-=0$, and hence that $[ 2 \partial_y K - K^*_{ab} K^{* ab} ]^+_-=0$.  The trace and trace-free parts of Eq. (\ref{bc1}) are then given by
\ba
\label{bcd}
\fpp(R) \left[ \partial_y R \right]^+_- &=& 0\;,\\
\label{bcb}
\fp (R) \left[ K^*_{ab} \right]^+_- &=& 0\;,\\
\label{bcc}
\left[ K \right]^+_- &=&0\;,
\ea
which, together with
\ba
\label{bca}
\left[\gamma_{ab} \right]^+_- &=& 0\;,\\
\label{bce}
\left[ R \right]^+_- &=& 0\;,
\ea
form the junction conditions in $f(R)$ theories in which $\fpp(R) \neq 0$. For further details the reader is referred to \cite{der}. 
\section{Matching to FLRW}
\label{sec:swiss}
One of the oldest ways of trying to construct inhomogeneous cosmological models is to join FLRW solutions, at some appropriate boundary, to the spherically symmetric spacetimes that are expected to exist around individual isolated objects.  This was famously achieved by Einstein and Straus for the case of the Schwarzschild solution and the Einstein-de Sitter universe \cite{ES}. It is also possible to join the Lema\^{i}tre-Tolman-Bondi solutions of Einstein's equations to FLRW at a spherical boundary \cite{Rib}.  These models are often referred to as `Swiss cheese', as this is what the global structure starts to look like if one can keep removing regions of the FLRW `cheese', and replacing it with either Schwarzschild or Lema\^{i}tre-Tolman-Bondi holes. The simplicity of this approach, and the degree to which it has influenced the development of inhomogeneous cosmology in general relativity, makes it a natural place to begin studying the relationship between weak-field systems and cosmology in $f(R)$ theories of gravity.

Here we will consider an FLRW geometry given by
\be
\label{FLRW}
ds^2 = -d\hat{t}^2 + a^2(\hat{t}) \left[ \frac{d\hat{r}^2}{1-k
    \hat{r}^2} + \hat{r}^2 d\hat{\theta}^2+ \hat{r}^2 \sin^2 \hat{\theta} d\hat{\phi}^2 \right],
\ee
and that is filled with a perfect fluid.  Within this spacetime we will excise a region interior to the sphere $\hat{r}=\hat{\Sigma}$, and replace it with a region of spacetime that is spherically symmetric, and that is well described by the weak-field geometry given in Eq. (\ref{Newtonian}).  In this case it is convenient to write the spatial metric in spherical polar coordinates, so that $\delta_{ij}dx^i dx^j = dr^2 + r^2 d\theta^2 + r^2 \sin^2 \theta^2 d\phi^2$.  We can then identify the angular coordinates in both regions, which we will refer to as Region I and Region II, respectively.

Without loss of generality, we consider the boundary to be comoving with the fluid. As there are no spatial gradients in Region I, the boundary must be static with respect to the hypersurfaces of homogeneity that exist in the FLRW geometry.  In Region II, however, the boundary is free to move in the radial direction.  The first fundamental form of the boundary, on either side, is then given by
\begin{eqnarray*}
\gamma^{I}_{ab} dx^a dx^b &=& -d\hat{t}^2 + a^2 \hat{\Sigma}^2 d
\Omega^2\;,\\
\gamma^{II}_{ab} dx^a dx^b &\simeq& -\left(1- 2 \phi - \dot{\Sigma}^2 \right)
dt^2 + (1+2 \psi) \Sigma^2 d\Omega^2\;,
\end{eqnarray*}
where the boundary is at $r=\Sigma$ in Region II, and where we have used the notation $\simeq$ to mean equal up to terms of post-Newtonian order (i.e. up to $O(\epsilon^3)$).  The junction condition (\ref{bca}) then gives the conditions
\begin{eqnarray}
\label{jc1}
(1+\psi) \Sigma &\simeq& a \hat{\Sigma}\;,\\
\label{jc2}
\frac{d\hat{t}}{dt} &\simeq& 1 - \phi-\frac{1}{2} \dot{\Sigma}^2\;.
\end{eqnarray}

Now let us consider the extrinsic curvature.  To calculate this we
need to know the normal to the boundary, which is given in each region by
\begin{eqnarray}
n^{I}_{\mu} &=& \frac{a
  \delta^{\hat{r}}_{\phantom{\hat{t}}\mu}}{\sqrt{1-k \hat{r}^2}}\;,\\
n^{II}_{\mu} &\simeq& \left( 1+\psi + \frac{1}{2} \dot{\Sigma}^2 \right) \delta^r_{\phantom{r} \mu} - \dot{\Sigma} \delta^t_{\phantom{t} \mu}.
\end{eqnarray}
The second fundamental form on the boundary is then given by
\be
\label{exttrans}
K_{ab} =\frac{\partial x^\mu}{\partial x^a} \frac{\partial
  x^\nu}{\partial x^b} n_{\mu ; \nu}\;,
\ee
which for the two spacetimes we are considering is
\begin{eqnarray}
K^{I}_{ab} dx^a dx^b &\simeq& r \left( 1- \frac{k r^2}{2 a^2} +
\frac{1}{\fp_0} U - \frac{\fpp_0}{2\fp_0} R \right) d \Omega^2,
\end{eqnarray}
and
\begin{eqnarray}
&&K^{II}_{ab} dx^a dx^b \\&\simeq& \left(\frac{1}{\fp_0} \hat{U}_{,r} +\frac{\fpp_0}{2\fp_0}
R_{,r}-\ddot{\Sigma} \right) d \hat{t}^2 
 \nonumber \\&& \nonumber + r \Big( 1- \frac{\fpp_0}{2\fp_0} R + \frac{1}{2} \dot{\Sigma}^2 - \frac{\fpp_0}{2\fp_0} r
R_{,r} +\frac{1}{\fp_0} U +\frac{r}{\fp_0} U_{,r} \Big) d \Omega^2,
\end{eqnarray}
where we have already used the junction conditions (\ref{jc1}) and (\ref{jc2}).  

The junction conditions (\ref{bcb}) and (\ref{bcc}) then give
\begin{eqnarray}
\label{sigdot1}
\frac{\dot{\Sigma}^2}{\Sigma^2} &\simeq& -\frac{2 U_{,r}\vert_{\Sigma}}{\fp_0 \Sigma} -
\frac{k \hat{\Sigma}^2}{\Sigma^2} +\frac{\fpp_0}{\fp_0} \frac{R_{,r}\vert_{\Sigma}}{\Sigma}\;,\\
\label{sigdot2}
\frac{\ddot{\Sigma}}{\Sigma} &\simeq& \frac{\hat{U}_{,r}\vert_{\Sigma}}{\fp_0 \Sigma} + \frac{\fpp_0}{2\fp_0} \frac{R_{,r}\vert_{\Sigma}}{\Sigma}.
\end{eqnarray}
These look very much like the Friedmann equations derived from Einstein's equations, with the terms containing the Newtonian potential $U$ acting like the matter terms, and with the term involving the spatial curvature $k$ playing its usual role.  Here, however, we also have two additional terms containing derivatives of the Ricci scalar, $R$. These extra terms can be seen to contain all of the new behaviour that one obtains by generalising the gravitational Lagrangian from $R$ to $f(R)$.

So far we have only applied the junction conditions that exist in Einstein's equations: That the first and second fundamental forms on the boundary must be continuous if we are to avoid a surface layer of matter. Let us now apply the additional junction condition (\ref{bcd}). The spacetime in Region I is homogeneous, so in this case we must have
\be
\partial_y R = \frac{\sqrt{1-k \hat{r}^2}}{a} R_{,\hat{r}} =0.
\ee
Applying the junction condition (\ref{bcd}) then gives
\be
R_{,r}\vert_{\Sigma} \simeq 0,
\ee
where we have used $k\hat{\Sigma}^2 \sim O(\epsilon^2)$, as can be seen from Eq. (\ref{sigdot1}). This means that the last terms on the right-hand side of both Eq. (\ref{sigdot1}) and Eq. (\ref{sigdot2}) must vanish at $O(\epsilon^2)$, so that we are left with exactly the same equations as in Einstein's theory (up to the presence of $\fp_0$ in the denominator of the terms involving $U$, which can be absorbed into constants, and the terms involving $f_0$ in Eqs. (\ref{PPNU}) and (\ref{PPNR}), which act like $\Lambda$.).

This treatment appears to show that the only Swiss cheese solutions that exist in $f(R)$ theories of gravity must either have FLRW regions that behave in the same way they do in Einstein's theory (possibly with $\Lambda$, and up to possible small corrections), or it must be the case that the spacetime within the excised spheres cannot be described using the weak-field geometry given in Eq. (\ref{Newtonian}), and explained in Section \ref{sec:weak}.
\section{Matching Without FLRW}
\label{sec:lattice}
In the previous section we tried to join a region described by a weak-field perturbations about Minkowski space to a FLRW geometry.  Here we take a different approach, and instead try and join together numerous different regions each of which is well described internally by a weak-field geometry of the form given in Eq. (\ref{Newtonian}).  Note that we do {\it not} require multiple regions to be well described by the same weak-field metric, but instead try and relate the different weak-field systems to each other by using the junction conditions discussed in Section \ref{sec:junction}.

We will proceed by taking a number of objects with the same mass and distributing them regularly in space.  We will then take the domain of each object to consist of all the points in space that are closest to that object.  What we mean by a `regular distribution' here is then that the domains of all objects in the spacetime should be identical, up to translations and rotations.  An example of this in $R^3$ is obtained by dividing the space up into a cubic lattice, and placing a mass at the center of each cube, the interior of which then acts as the domain of the mass at its center.

These situations have been considered within the context of Einstein's equations by a number of authors.  Approximate matching schemes have been developed and studied for joining together Schwarzschild solutions in the spirit of the Wigner-Seitz construction from solid state physics \cite{LW1,CF1,CF2,CF3}.  Perturbative treatments have also been attempted \cite{clif,julien1,julien2}, as well as an exact treatment of the initial value problem \cite{CRT}, and numerical studies \cite{num1,num2}.  This type of model seems ideally suited to a study of the relationship between weak-field systems and cosmology, as they allow one to perform a bottom-up construction of a cosmological model from the weak-field systems themselves. They also do not require the existence of FLRW geometry at the boundaries between regions, and so one is allowed to move away from some of the restrictions of the Swiss cheese models discussed in Section \ref{sec:swiss}.

From the symmetry of the situation the junction condition (\ref{bca}) is automatically satisfied.  We then need to consider the conditions (\ref{bcb}) and (\ref{bcc}).  To do this we need to know the extrinsic curvature of the boundary of each domain.  If the unit normal to this hypersurface is given by $n_{\mu} = (n_t ; n_i)$ then, in the geometry (\ref{Newtonian}) this is given to lowest non-trivial order by \cite{clif}
\ba
K_{\mu \nu} dx^{\mu} dx^{\nu} &\simeq& \left(n_{t,t} + n_{i} \phi_{,i} \right) dt^2 + \left(n_{i,t}+n_{t,i} \right) dx^i dt \nonumber \\ \nonumber
&&+\left(n_{i,j} -2 \psi_{i} n_j + \delta_{i,j} \delta^{kl} \psi_{,k} n_l \right) dx^i dx^j\;.
\label{K1}
\ea
The magnitude of $n_t$ can be seen to be of $O(\epsilon)$ here, while $n_i$ is $O(1)$, as the condition $u^{\mu} n_{\mu}=0$, where $u^{\mu}$ is the 4-velocity of the boundary, gives $n_t = -n_i u^i$, and $u^i$ is $O(\epsilon)$.

We must now transform $K_{\mu \nu}$ into $K_{a b}$ using Eq. (\ref{exttrans}). To do this it is convenient to pick out a direction, which we call $z$, that is normal to the boundary at the point where it intersects a straight line that joins the positions of two neighbouring masses.  The remaining two spatial directions, which we will use together with $t$ as the intrinsic coordinates on the boundary, will then be written using indices $A, B, C$ {\it etc.}.  The position of the boundary itself will be given by $z=Z(t,x^A)$.  The extrinsic curvature of the boundary is then \cite{clif}
\begin{widetext}
\ba
K_{ab} dx^a dx^b &\simeq& - n_z \Bigg[ \left(\ddot{Z} -\phi_{,z} +
  Z_{\vert A} \phi_{,A} \right) dt^2
+ \left( (Z_{\vert A})\dot{\;} + (\dot{Z})_{\vert A} \right) dx^A dt
\nonumber \\ && \qquad \qquad \qquad
+ \Big( Z_{\vert AB} - (\psi_{,z}-\delta^{CD} Z_{\vert C} \psi_{\vert
    D}) (\delta_{AB} + Z_{\vert A} Z_{\vert B}) \Big) dx^A dx^B
\Bigg],
\label{K3}
\ea
where $\dot{\;} \equiv u^{\mu} \partial_{\mu} = \partial_t + Z_{,t} \partial_z$ and $\:_{\vert A} \equiv m^{\mu} \partial_{\mu}= \partial_{A} + Z_{,A} \partial_z$.  Here $m^{\mu}$ is a space-like unit vector in the boundary, and we have taken $u^A=0=m^t$. All quantities in Eq. (\ref{K3}) should be taken to be evaluated at the boundary.
\end{widetext}

It can now be seen that in order to simultaneously satisfy Eqs. (\ref{bcb}) and (\ref{bcc}), as well the reflective symmetry about the boundary that is required by our regular distribution of masses, we must have $K_{ab}=0$. From Eq. (\ref{K3}) this can be seen to correspond to \cite{clif}:
\ba
\label{static1}
\frac{\ddot{Z}}{\sqrt{1+(Z_{\vert A})^2}} &\simeq& n \cdot \nabla \phi\;,\\
\frac{Z_{\vert AB}}{\sqrt{1+(Z_{\vert A})^2}} &\simeq& (\delta_{AB} +Z_{\vert A} Z_{\vert B}) (n \cdot \nabla \psi), \label{static2}
\ea
as well as $(Z_{\vert A})\dot{\;} \simeq \big( \dot{Z} \big)_{\vert A} \simeq 0$.
These equations govern the motion and shape of the boundary in these highly symmetric configurations.

Finally, let us apply the junction conditions (\ref{bcd}) and (\ref{bce}). The latter of these is again automatically satisfied from the symmetry about the boundary.  The former, however, gives $n \cdot \nabla R=0$, which means that Eqs. (\ref{static1}) and (\ref{static2}) become
\ba
\label{static3}
\frac{\ddot{Z}}{\sqrt{1+(Z_{\vert A})^2}} &\simeq& \frac{1}{\fp_0} n \cdot \nabla \hat{U}\;,\\
\frac{Z_{\vert AB}}{\sqrt{1+(Z_{\vert A})^2}} &\simeq& (\delta_{AB} +Z_{\vert A} Z_{\vert B}) \frac{1}{\fp_0} n \cdot \nabla U. \label{static4}
\ea
These are exactly the same expressions that are found using Einstein's equations (up to the factors of $1/f^{\prime}_0$, which can again be absorbed into constants, and terms involving $f_0$, which again act like $\Lambda$).  The solutions to Eqs. (\ref{static3}) and (\ref{static4}) are known to be either decelerating, or correspond to Minkowski space, unless $f_0\neq 0$.

Once again, we find that for the weak-field regions to exist we must have cosmological behaviour that is the same as in Einstein's theory (up to possible small corrections). Therefore, within this approach, the large-scale behaviour must reduce to that expected from Einstein's equations, or one must relinquish the usual weak-field description of perturbed Minkowski space around astrophysical objects.
\section{Matching Exact Solutions}
\label{sec:exact}
We have so far considered joining weak-field geometries to either FLRW, or to each other, in theories in which $f(R)$ is an analytic function.  This has shown that acceleration in the resulting cosmological model cannot occur in any new ways if the junction conditions given in Section \ref{sec:junction} are to be obeyed. One must therefore either allow for gravitational fields to be rapidly varying, or give up on a description of the regions around astrophysical objects as small fluctuations about Minkowski space.  The latter of these two possibilities suggests that it may be useful for us to consider exact solutions.

Unfortunately, the complexity of the field equations (\ref{FE}) make exact solutions difficult to find. However we know that {\it for all functions $f(R)$ which are of class $C^3$ at $R=0$ and $f(0)=0$ while $f'(0)\ne 0$, the Schwarzschild solution is the only vacuum solution with vanishing Ricci scalar} \cite{anne}. It therefore seems natural to try and match a spherical region with Schwarzschild geometry to an exterior FLRW spacetime. In the context of Einstein's theory this corresponds to the well-known Einstein-Straus approach described earlier \cite{ES}. Furthermore, if we restrict our considerations to $f(R)=R^{1+\delta}$ then there are two known exact solutions (other than the vacuum solutions of Einstein's equations, that is, which are also solutions of these theories).  A static spherically symmetric vacuum solution with non-trivial asymptotics was found in \cite{power}, and a time-dependent spherically symmetric vacuum solution was found in \cite{sph}.  In what follows, we will also try and join these two solutions to FLRW geometries.
\subsection{An Einstein-Straus-like Construction}
The constructions we consider here consist of  point-like masses at the centre of otherwise empty spherical regions, whose geometry is described by the Schwarzschild metric, and that are embedded in FLRW geometry at appropriate boundaries. Such constructions were originally considered by Einstein and Straus \cite{ES}, and were introduced to address the question of whether or not the expansion of the universe can affect local mechanical phenomena, such as planetary orbits. Since the spacetime near the central mass is Schwarzschild, the planetary orbits are given by the usual time-like geodesics of this geometry, and the cosmic expansion does not affect them. Let us now investigate whether such a construction can be performed in $f(R)$ gravity. 

We begin by writing the Schwarzschild solution as
\be
\label{Sch}
ds^2 = -A(r) dt^2 + \frac{dr^2}{A(r)}+r^2 \left(d \theta^2 +\sin^2 \theta d \phi^2 \right)\;,
\ee
where 
\be
A(r)=\left(1-\frac{2M}{r}\right)\;.
\ee
Let us now try and embed this solution in an FLRW geometry, as specified in Eq. (\ref{FLRW}).
To do this, consider a boundary at $\hat{r}=\hat{\Sigma}$ in the FLRW spacetime and $r=\Sigma$ in the Schwarzschild solution.  
The first fundamental form on the boundary is then given in the vacuum region by
\be
\gamma_{ab} dx^a dx^b = -\left(A-\frac{\dot{\Sigma}^2}{A} \right)
dt^2+\Sigma^2 d\Omega^2
\ee
and in the FLRW region by
\be
\gamma_{ab} dx^a dx^b = -d\hat{t}^2+a^2(\hat{t})\hat{\Sigma}^2d\Omega^2,
\ee
where we have identified angular coordinates in the two different regions at the boundary and where $d\Omega^2 = d\theta^2+\sin^2 \theta d\phi^2$.
The junction condition (\ref{bca}) then gives
\ba\label{f1}
\Sigma& = &a(\hat{t}) \hat{\Sigma}\;,
\ea
\ba\label{f2}
\frac{d\hat{t}}{dt} &=& \sqrt{A-\frac{\dot{\Sigma}^2}{A}}.
\end{eqnarray}
To calculate the second fundamental form we need the space-like unit
vector normal to the boundary.  In the vacuum region this is given by
\be
n_{\mu} = \frac{\sqrt{A}}{\sqrt{A^2-\dot{\Sigma}^2}} \left( - \dot{\Sigma},1,0,0 \right),
\ee
while in the FLRW region it is
\be
n_{\mu} = \frac{a(\hat{t}) \delta^{r}_{\phantom{r} \mu}}{\sqrt{1-k \hat{\Sigma}^2}}\;.
\ee
The second fundamental form on the FLRW side of the boundary is then
\be
K_{ab} dx^a dx^b = a(\hat{t}) \hat{\Sigma} \sqrt{1-k\hat{\Sigma}^2} d \Omega^2,
\ee
while on the vacuum side of the boundary it is given by
\begin{eqnarray}
K_{tt} &=& \frac{3 AA_{,r} \dot{\Sigma}^2-A^3A_{,r} -2 A^2 \ddot{\Sigma}}{2 \sqrt{A} (A^2-\dot{\Sigma}^2)^{3/2}}\;,\\
K_{\theta \theta} &=& \sqrt{\frac{\Sigma^2 A^3}{(A^2-\dot{\Sigma}^2)}}\;,
\end{eqnarray}
where all quantities should be evaluated at the boundary.
Matching $K_{\theta \theta}^{\pm}$ at the boundary we obtain 
\ba
\dot{\Sigma}^2 &=& A^2\left[1 -\frac{A}{(1-k \Sigma^2/a^2)}\right]\;.
\ea
Writing the above equation in the coordinates $(\hat{t},\hat{r},\theta,\phi)$, and using Eqns. (\ref{f1}), (\ref{f2}) together with the 
form of the function $A(r)$, we find
\be
\hat{\Sigma}^3a(\hat{t})\left[ k+\left( \frac{da(\hat{t})}{d\hat{t}}\right)^2\right]=2M\;.
\label{mass}
\ee
The left-hand side of the above equation is the usual definition of the Cahil-Macvitte function in FLRW spacetimes. 


Differentiating Eqn. (\ref{mass}) with respect to $\hat{t}$ gives $G^1_{\phantom{1}1}=0$.  This implies that the total pressure (standard matter and curvature fluid) must vanish on the boundary, but as the pressure in the FLRW region is homogeneous, this means that the total pressure should vanish throughout 
the FLRW region. In this case, equating the time component of the extrinsic curvature will not give any new information.

If we now impose the requirement that $R$ should be the same on either side of the boundary, from Eq. (\ref{bce}), then we must have 
\be
6\left[\frac{1}{a(\hat{t})}\frac{d^2a(\hat{t})}{d\hat{t}^2}+\frac{1}{a(\hat{t})^2}\left(\frac{da(\hat{t})}{d\hat{t}}\right)^2+\frac{k}{a(\hat{t})^2}
\right]=0\;.
\ee
The above equation combined with the condition of vanishing total pressure, then implies vanishing total density (curvature fluid and standard matter) in the FLRW region. Whatismore, putting $R=0$ in Eq. (\ref{FE}) shows that the effective energy-momentum tensor of the curvature fluid must be proportional to $g_{\mu\nu}$. It then follows that the energy-momentum tensor of standard matter, $T_{\mu \nu}^m$, must also be proportional to $g_{\mu \nu}$, and so can only be a vacuum energy. It also follows that the FLRW region can only be Minkowski spacetime (in Milne coordinates, if $k=-1$). Finally, from Eq. (\ref{bcd}) we see that the normal gradients automatically match identically, as $R=0$ on both sides. 

We note that the situation remains the same if instead of a Schwarzschild interior we have a Schwarzschild-de Sitter, or anti-de Sitter, interior. In these cases the interior region has a constant, non-zero Ricci scalar.  As $R$ must be matched across the boundary, the FLRW region must also have a constant Ricci scalar, and from Eq. (\ref{FE}) it can be easily seen that the effective energy-momentum tensor of the curvature fluid must be proportional to $g_{\mu\nu}$. Furthermore, matching the second fundamental form now gives $G^1_{\phantom{1}1}=\rm{constant}$ in the FLRW region, which implies that the total pressure must be constant. Taken together, these two conditions imply that the total energy density should also be constant, and that the energy-momentum tensor of matter in the FLRW region must have $T_{\mu \nu}^m \propto g_{\mu \nu}$, which is nothing other than vacuum energy. The only solution in this case is therefore a spacetime that is de Sitter everywhere.

It is a curious result that the Schwarzschild solution cannot be embedded in any FLRW spacetime (other than the trivial case of Minkowski space) in $f(R)$ theories of gravity, unless the theory is linear in $R$. However, this conclusion is natural from the junction conditions. This is because the conditions that the Ricci scalar and its first derivative should match across the boundary make the non-trivial $f(R)$ theories qualitatively different from general relativity, where $R$ can be discontinuous. If a spherically symmetric object is joined to a FLRW geometry in $f(R)$ theories, then one must expect an evolution of the boundary values of $R$ and $\dot{R}$, which is something that pure Schwarzschild or Schwarzschild-de Sitter solutions cannot satisfy. Hence, in the following sections, we will explore some other exact non-GR solutions in $f(R)$ gravity, in order to check whether Einstein-Straus-like constructions are possible with them.

\subsection{A Static Solution in $R^n$ Gravity}
An exact static, spherically symmetric vacuum solution of $f(R)=R^{1+\delta}$ is given by \cite{power}
\be
\label{AB}
ds^2 = -A(r) dt^2 + \frac{dr^2}{B(r)}+r^2 \left(d \theta^2 +\sin^2 \theta d \phi^2 \right),
\ee
where
\ba
\nonumber
A(r) &=& r^{\frac{2\delta (1+2 \delta)}{(1-\delta)}} +
\frac{c_1}{r^{\frac{(1-4 \delta)}{(1-\delta)}}}\;,\\
\nonumber
B(r) &=& \frac{(1-\delta)^2}{(1-2 \delta+4 \delta^2)(1-2\delta-2
  \delta^2)} \left( 1+
\frac{c_1}{r^{\frac{(1-2\delta+4\delta^2)}{(1-\delta)}}} \right).
\ea
The Ricci scalar for this solution is
\be
R=-\frac{6 \delta(1+\delta)}{(1-2 \delta-2\delta^2) a^2 r^2}.
\ee
We will now try and embed this solution in an FLRW geometry.
To do this, consider a boundary at $r=\Sigma$ in the vacuum region.  The first fundamental form on the boundary is then given in the vacuum region by
\be
\gamma_{ab} dx^a dx^b = -\left(A-\frac{\dot{\Sigma}^2}{B} \right)
dt^2+\Sigma^2 d\Omega^2.
\ee
Matching the first fundamental forms then gives
\ba
\Sigma &=& a(\hat{t}) \hat{\Sigma}\\
\frac{d\hat{t}}{dt} &=& \sqrt{A-\frac{\dot{\Sigma}^2}{B}}.
\end{eqnarray}
In the vacuum region the spacelike unit vector normal to the boundary is given by
\be
n_{\mu} = \frac{\sqrt{A}}{\sqrt{AB-\dot{\Sigma}^2}} \left( - \dot{\Sigma},1,0,0 \right),
\ee
The second fundamental form of the vacuum side is 
\begin{eqnarray}
K_{tt} &=& \frac{2 BA_{,r} \dot{\Sigma}^2+A B_{,r} \dot{\Sigma}^2-A B^2 A_{,r} -2 AB \ddot{\Sigma}}{2 \sqrt{A} (AB-\dot{\Sigma}^2)^{3/2}}\;,\\
K_{\theta \theta} &=& \sqrt{\frac{\Sigma^2 B^2 A}{(AB-\dot{\Sigma}^2)}}\;,
\end{eqnarray}
where all quantities should be evaluated at the boundary.

The junction conditions (\ref{bcb}) and (\ref{bcc}) are then satisfied if
\begin{eqnarray}
\dot{\Sigma}^2 &=& AB \left[1 -\frac{B}{(1-k \Sigma^2/a^2)}\right]\\
\ddot{\Sigma} &=& \frac{(A_{,r}B+B_{,r}A)}{2} -\frac{B (2
  A_{,r}B+B_{,r}A)}{2 (1-k \Sigma^2/a^2)}.
\end{eqnarray}
Consistency of these equations requires
\be
\frac{(A B_{,r} - A_{,r}B)}{(1-k r^2)} =0.
\ee
Substitution from Eq.~(\ref{AB}) shows that this can be achieved only if $\delta=0$ or $-1/2$.

If we now impose the requirement that $R$ should be the same on either side of
the boundary, from Eq. (\ref{bce}), then we get
\ba
\label{Rstatic}
\frac{1}{a(\hat{t})}\frac{d^2a(\hat{t})}{d\hat{t}^2}+\frac{1}{a(\hat{t})^2}\left(\frac{da(\hat{t})}{d\hat{t}}\right)+\frac{k}{a(\hat{t})^2}
&=&\nonumber\\
-\frac{\delta(1+\delta)}{(1-2 \delta-2\delta^2) a(\hat{t})^2 \hat{\Sigma}^2}.&&
\ea
This strongly constrains the allowed form of $a(t)$.  Finally, from Eq. (\ref{bcd}), we find that we must have 
\be
\frac{\sqrt{A} B R_{,r}}{\sqrt{AB-\dot{\Sigma}^2}} = 0,
\ee
as there are no spatial gradients in the FLRW region. This means that we must also require $R_{,r}=0$ at the boundary in the vacuum region. This is only satisfied if $\delta =0$ or $-1$, as can be seen from the right-hand side of Eq. (\ref{Rstatic}).

It is therefore the case that the junction conditions from Section \ref{sec:junction} can only be satisfied if $\delta =0$, in which case the field equations (\ref{FE}) simply reduce to Einstein's equations.  In this case the vacuum solution given in Eq. (\ref{AB}) reduces to the Schwarzschild solution, and Eq. (\ref{Rstatic}) no longer needs to be satisfied as $\fpp=0$, and the right-hand side of Eq. (\ref{Rjunc}) vanishes automatically.  The vacuum solution (\ref{AB}) cannot, therefore, be used to model the gravitational field of an astrophysical object embedded in an FLRW universe in these theories, unless $f(R)$ is linear in $R$.  This is despite the fact that this solution is the asymptotic attractor of all spherically symmetric, static, vacuum solutions of theories with $f(R)=R^{1+\delta}$ \cite{power}, suggesting that the spacetime around astrophysical objects that are embedded in FLRW should be time dependent.
\subsection{A Non-Static Solution in $R^n$ Gravity}
An exact solution for time-dependent, spherically symmetric vacuum situations in $f(R)=R^{1+\delta}$ theories is given by \cite{sph}:
\be
\label{ns}
ds^2=-A(r) dt^2 +q^2(t) B(r) \left( dr^2 +r^2 d\Omega^2 \right)\;,
\ee
where $q(t) = t^{\frac{\delta (1+2 \delta)}{(1-\delta)}}$, and
\ba
\nonumber
A(r) &=& \left[ \frac{1-\frac{c_2}{r}}{1+\frac{c_2}{r}} \right]^{2/\sigma}\;,\\
\nonumber
B(r) &=& \left( 1+ \frac{c_2}{r} \right)^4 A^{\sigma+2 \delta-1}\;,
\ea
where $\sigma^2 = 1-2 \delta +4 \delta^2$. The Ricci scalar in this case is given by
\be
R = - \frac{6 \delta (1+\delta) (1+2 \delta) (1-4\delta)}{(1-\delta)^2 t^2 A}\;.
\ee
Again, to match this solution with a FLRW exterior, consider a boundary at $r=\Sigma$ in this solution. The first fundamental form on the boundary is then given in the vacuum region by
\be
\gamma_{ab} dx^a dx^b = -\left(A-q^2B\dot{\Sigma}^2\right)
dt^2+q^2B\Sigma^2 d\Omega^2\;.
\ee
Matching the first fundamental forms then gives
\ba
\label{first1ff}
q\sqrt{B}\Sigma &=& a(\hat{t}) \hat{\Sigma}\\
\frac{d\hat{t}}{dt} &=& \sqrt{A-q^2B\dot{\Sigma}^2}.
\label{second1ff}
\end{eqnarray}
and the unit vectors tangent and normal to the boundary are given by
\begin{eqnarray}
u^{\mu} &=& \frac{1}{\sqrt{A-B q^2 \dot{\Sigma}^2}} (1,\dot{\Sigma},0,0)\;,\\
n_{\mu} &=& \frac{\sqrt{AB} q}{\sqrt{A-B q^2 \dot{\Sigma}^2}} (-\dot{\Sigma},1,0,0).
\end{eqnarray}
Calculating the second fundamental form for the matching surface we get
\begin{widetext}
\ba
K_{tt}&=&\frac{2\dot{q}q(B^2q^2\dot{\Sigma}^3-2AB\dot{\Sigma})+q^2\dot{\Sigma}^2(2A_rB-B_rA)-2q^2AB\ddot{\Sigma}-A A_r}{2\sqrt{AB}q(A-B q^2 \dot{\Sigma}^2)^{3/2}}\;,\\
K_{\theta\theta}&=&\frac{q\Sigma(AB_r\Sigma+2AB+2\dot{q}\dot{\Sigma}B^2q\Sigma)}{2\sqrt{AB}\sqrt{A-B q^2 \dot{\Sigma}^2}}.
\label{kab}
\ea
Matching the Ricci scalar then gives 
\ba
\label{Rdynamic}
\frac{1}{a(\hat{t})}\frac{d^2a(\hat{t})}{d\hat{t}^2}+\frac{1}{a(\hat{t})^2}\left(\frac{da(\hat{t})}{d\hat{t}}\right)^2+\frac{k}{a(\hat{t})^2}
  =- \frac{\delta (1+\delta) (1+2 \delta) (1-4\delta)}{(1-\delta)^2 t^2 A(\Sigma)}.
\ea
Finally, the boundary condition $n \cdot \nabla R=0$, gives
\be
\label{nR}
\dot{\Sigma} = - \frac{2 c_2}{\sigma \Sigma^2 \left(1-\frac{c_2}{\Sigma}\right)^3
\left(1+\frac{c_2}{\Sigma}\right)^3} \left( \frac{1-\frac{c_2}{\Sigma}}{1+\frac{c_2}{\Sigma}}
\right)^{\frac{4(1-\delta)}{\sigma}} t^{\frac{(1-3 \delta-4 \delta^2)}{(1-\delta)}},
\ee
unless $\delta=1/4$, $0$, $-1$ or $-1/2$, in which case $n \cdot \nabla R=0$ automatically. We can now construct an algebraic constraint for $\Sigma$ by equating $K_{tt}$ on either side of the boundary and using Eqs. (\ref{first1ff}) and (\ref{nR}) to remove $a(\hat{t})$ and $\dot{\Sigma}$. This gives
\be
\frac{q}{A^{\frac{1-2 \delta}{2}}} \left[\frac{ \left(\Sigma^2+c_2^2\right) (1-\delta) k -2 c_2 \sigma^2 \Sigma}{(1-\delta) \sqrt{\sigma^2 - 4 c_2^2 t^{\frac{2 (1-2 \delta-2\delta^2)}{(1-\delta)}} \Sigma^{8(1-\delta)/\sigma} (\Sigma^2-c_2^2)^2 A^{2(1-\delta)}}} - \sqrt{1-k\hat{\Sigma}^2} \left(\Sigma^2-c_2^2 \right) \right]=0.
\ee
\end{widetext}
This equation must be satisfied at all times, but is clearly very difficult to solve for $\Sigma$ directly. We can, however, perform a series expansion in $c_2$. To zero order we then have the constraint
\be
\sqrt{1-k \hat{\Sigma}^2} = 1 + O(c_2),
\ee
so that $k \simeq 0$. This says that the FLRW geometry that we are embedding within must be close to spatially flat. Using this in the first order equation then gives
\be
\frac{\sigma}{(1-\delta)} c_2= 0+O(c_2^2),
\ee
so that the only possible solutions would appear to require either $\sigma=0+O(c_2)$, or $c_2=0$. The first of these possibilities requires $\delta$ to be complex, which we are not interested in here, and the second is the requirement that the central mass vanishes. The matching of this latter situation to FLRW is trivial, as the geometry in Eq. (\ref{ns}) can itself be seen to reduce to FLRW when $c_2 \rightarrow 0$. Once again we therefore appear to be unable to match solutions to FLRW, except when $\delta=0$, or when the entire spacetime is FLRW anyway.


The anomalous cases that remain are those in which $\delta=1/4$, $-1$ or $-1/2$, as in these cases Eq. (\ref{nR}) can no longer be used. Of these $\delta=-1$ seems problematic as it corresponds to a Lagrangian density $\mathcal{L}=$constant, which can hardly be said to be a Lagrangian for gravity at all. The cases $\delta=-1/2$ and $\delta=1/4$ also seem problematic, as in these cases the field equations (\ref{FE}) contain terms that are ill-defined, with both numerator and denominator reducing to zero. 
In all of these cases the Ricci scalar must vanish, so the only exterior FLRW geometry that one could match to would have to be Milne anyway.  We do not, therefore, consider them to be of any interest for our current purposes.


We therefore find that even for this non-trivial non-static solution, a matching with a FLRW exterior is not possible. This is true even though the solution itself approaches FLRW asymptotically.

\section{Discussion}
\label{sec:conclude}
The idea that the late-time acceleration of the universe could be explained by modifications of the Einstein-Hilbert action has recently attracted considerable interest, but a complete understanding of the consequences of such a radical shift away from the standard approach to cosmology is still far from complete. Here we attempt to add to this understanding by considering the construction of cosmological models that contain massive astrophysical bodies within the context of $f(R)$ theories of gravity. Such constructions are key to understanding the effect cosmological expansion has on the gravitational fields of astrophysical bodies, as well as describing the large-scale expansion that emerges in a universe with large density contrasts.

After a discussion of the junction conditions that need to be satisfied when matching together different solutions in $f(R)$ theories, a number of attempts were made to construct inhomogeneous cosmological models by matching different regions of spacetime. This was done both for theories with general analytic functions $f(R)$ and for non-analytic theories with $f(R)=R^n$.  In all cases studied, it was found that it is impossible to satisfy the required junction conditions without the large-scale behaviour reducing to what is found from Einstein's equations with a cosmological constant.  For theories with analytic $f(R)$ this suggests that the usual treatment of weak-field systems as perturbations around Minkowski space may not be compatible with late-time acceleration that is driven by anything other an effective cosmological constant given by $f(0)$.  For theories with $f(R)=R^n$,  we found that a number of well-known spherically symmetric vacuum solutions could not be matched to an expanding FLRW background, including the well-known Einstein-Straus-like embeddings of the Schwarzschild exterior solution in FLRW spacetimes. 

The absence of these constructions represents a crucial difference between $f(R)$ theories and scalar-tensor theories of gravity. In the latter it is already known that Einstein-Straus-like embeddings are indeed possible, both in cosmological and astrophysical gravitational collapse scenarios (see for example \cite{harada}). This is true despite the extra junction conditions that are required in scalar-tensor theories, where the scalar field and its normal derivative must be matched at the boundary. These two conditions may initially seem quite similar to the extra conditions required in $f(R)$ gravity ({\it i.e.} matching the Ricci scalar and its normal derivative). However, it turns out that the conditions in $f(R)$ theories are much more restrictive, and give much stronger constraints on the spacetimes allowed on either side of the boundary. This is due to $R$ taking a very specific form once an ansatz has been made for the metric (by specifying it should be give by Eqs. (\ref{Newtonian}) or (\ref{Sch}), for example), which is in general not true for scalar-tensor theories.
 
These results are quite different to what is suggested by using linear perturbation theory around an FLRW background in $f(R)$ theories. In that case there seems to be little impediment to including large density contrasts by allowing $\delta \rho$ to become large, while $\phi$ and $\psi$ are required to stay small. This difference could indicate that while the weak-field solutions we have considered here are problematic, there may be ways of obtaining useful (approximate) spacetime geometries from the perturbed FLRW approach. This would, in fact, appear to be quite similar to the approach that is taken in \cite{olmo}, where the expansion of $f(R)$ is performed around a time-dependent, but spatially homogeneous and isotropic background geometry with $R=R_0(t)$. In this case small regions of spacetime can still be approximated as being close to Minkowski space, but the emergence of cosmological evolution on large scales cannot be studied in the same way, as it is, at least to some degree, being assumed from the outset. This does not in any way diminish the potential validity of such an approach, but it does appear to require knowledge about the geometry of the entire observable universe in order to model the spacetime around a single astrophysical object (it would also appear to require a re-think of the current framework for interpreting precision tests of gravity). Alternatively, it may be the case that the difference between the bottom-up constructions attempted here, and the top-down construction of perturbed FLRW, could be indicating that cosmological back-reaction is large in $f(R)$ theories. This is certainly plausible, and should probably be expected when ``screening mechanisms'' such as the chameleon effect come into play.

As a final thought, it would be interesting to study the physical consequences of `{\it jumps}' in the Ricci scalar and/or in the normal derivative of the Ricci scalar across the boundary.  As is well known from the Israel junction conditions, a jump in the second fundamental form gives rise to surface stress-energy and surface tension on the matching surface that can, for example, be used to stabilize gravitational vacuum condensate stars ~\cite{mazur}. In a similar way, it is plausible that relaxing the extra matching conditions in $f(R)$ theories could give rise to surface terms that might be of physical interest. This has been studied in the context of brane-world cosmology in \cite{der}.

\begin{center}
{\bf ACKNOWLEDGEMENTS}
\end{center}
We thank A. Coley and R. Tavakol for helpful discussions. TC acknowledges the support of the STFC, and is grateful to the University of Cape Town for hospitality while some of this work was carried out. The National Research Foundation (South Africa) is acknowledged for financial support. RG is supported by the Claude Leon Foundation, and AMN is supported by the National Astrophysics and Space Science Program (South Africa).


\end{document}